\documentclass[11pt,letterpaper]{article}

\usepackage[margin=1in]{geometry}
\usepackage[T1]{fontenc}
\usepackage{lmodern}
\usepackage{microtype}
\usepackage{amsmath}
\usepackage{amssymb}
\usepackage{booktabs}
\usepackage{xcolor}
\usepackage{tikz}
\usepackage{enumitem}
\usepackage{hyperref}

\definecolor{xtracePurple}{HTML}{6F2DBD}
\definecolor{xtraceViolet}{HTML}{A855F7}
\definecolor{xtraceInk}{HTML}{1F1235}

\hypersetup{
  pdftitle={Batched Paillier-Based Hamming-Distance Computation over Binary Embeddings},
  pdfauthor={Yavor Litchev, Liwen Ouyang},
  pdfsubject={CPU and GPU implementation study of encrypted Hamming-distance computation},
  pdfkeywords={Paillier encryption, Hamming distance, binary embeddings, GPU computing},
  pdfdisplaydoctitle=true,
  colorlinks=true,
  linkcolor=xtracePurple,
  citecolor=xtracePurple,
  urlcolor=xtracePurple
}

\title{Batched Paillier-Based Hamming-Distance\\
Computation over Binary Embeddings}
\author{
  Yavor Litchev\\
  \small\href{mailto:ylitchev@stanford.edu}{\texttt{ylitchev@stanford.edu}}
  \and
  Liwen Ouyang\\
  \small\href{mailto:liwen@xtrace.ai}{\texttt{liwen@xtrace.ai}}
}
\date{}

\DeclareMathOperator{\Ham}{Ham}
\DeclareMathOperator{\popcount}{popcount}
\DeclareMathOperator{\lcm}{lcm}
\DeclareMathOperator{\ord}{ord}
\newcommand{\BenchmarkRepeats}{3}
\newcommand{\BaselineCPUEncryptMs}{10.5320}
\newcommand{\BaselineCPUEncryptRate}{94.9}

\newcommand{\BaselineCPUEncryptMinMs}{10.4244}
\newcommand{\BaselineCPUEncryptMaxMs}{10.5369}
\newcommand{\BaselineCPUDecodeMs}{6.9328}
\newcommand{\BaselineCPUDecodeRate}{144.2}

\newcommand{\BaselineCPUDecodeMinMs}{6.9221}
\newcommand{\BaselineCPUDecodeMaxMs}{6.9472}
\newcommand{\LookupCPUEncryptMs}{0.6547}
\newcommand{\LookupCPUEncryptRate}{1,527.3}

\newcommand{\LookupCPUEncryptSpeedup}{16.1}
\newcommand{\LookupCPUEncryptMinMs}{0.6460}
\newcommand{\LookupCPUEncryptMaxMs}{0.6564}
\newcommand{\LookupCPUDecodeMs}{1.0678}
\newcommand{\LookupCPUDecodeRate}{936.5}

\newcommand{\LookupCPUDecodeSpeedup}{6.5}
\newcommand{\LookupCPUDecodeMinMs}{1.0673}
\newcommand{\LookupCPUDecodeMaxMs}{1.0759}
\newcommand{\BaselineGPUEncryptMs}{0.2652}
\newcommand{\BaselineGPUEncryptRate}{3,770.0}

\newcommand{\BaselineGPUEncryptSpeedup}{39.7}
\newcommand{\BaselineGPUEncryptMinMs}{0.2649}
\newcommand{\BaselineGPUEncryptMaxMs}{0.2702}
\newcommand{\BaselineGPUDecodeMs}{0.2235}
\newcommand{\BaselineGPUDecodeRate}{4,474.0}

\newcommand{\BaselineGPUDecodeSpeedup}{31.0}
\newcommand{\BaselineGPUDecodeMinMs}{0.2235}
\newcommand{\BaselineGPUDecodeMaxMs}{0.2316}
\newcommand{\LookupGPUEncryptMs}{0.0232}
\newcommand{\LookupGPUEncryptRate}{43,090.8}
\newcommand{\LookupGPUEncryptRateRounded}{43,091}
\newcommand{\LookupGPUEncryptSpeedup}{453.8}
\newcommand{\LookupGPUEncryptMinMs}{0.0231}
\newcommand{\LookupGPUEncryptMaxMs}{0.0235}
\newcommand{\LookupGPUDecodeMs}{0.0345}
\newcommand{\LookupGPUDecodeRate}{28,983.5}
\newcommand{\LookupGPUDecodeRateRounded}{28,983}
\newcommand{\LookupGPUDecodeSpeedup}{200.9}
\newcommand{\LookupGPUDecodeMinMs}{0.0344}
\newcommand{\LookupGPUDecodeMaxMs}{0.0351}
\newcommand{\LookupGPUOverCPUEncrypt}{28.2}
\newcommand{\LookupGPUOverCPUDecode}{30.9}

\begin{document}

\maketitle

\begin{abstract}
Additively homomorphic encryption supports outsourced computation on encrypted binary embeddings, but large-integer arithmetic and data movement can limit throughput. We describe a Paillier-based client that combines a carry-separated binary encoding, table-based encryption, reduced-exponent decryption, CUDA/CGBN arithmetic, persistent device state, and batched retrieval integration. We establish the encoding's correctness and characterize four CPU and GPU client configurations. The lookup configuration uses a 280-bit exponent-size parameter. Across \BenchmarkRepeats{} warm-state trials on batches of 10,000 random 512-bit embeddings, the lookup GPU configuration achieved median-batch throughputs of \LookupGPUEncryptRateRounded{} encryptions/s and \LookupGPUDecodeRateRounded{} Hamming-distance decodes/s. Its amortized costs were \LookupGPUEncryptMs\,ms and \LookupGPUDecodeMs\,ms per vector, corresponding to factors of $\LookupGPUEncryptSpeedup$ and $\LookupGPUDecodeSpeedup$ relative to the measured CPU baseline. These implementation-specific results demonstrate the throughput benefits of combining cryptographic precomputation, batched accelerator execution, and persistent runtime state. The study distinguishes warm-batch performance from isolated-request latency and identifies the remaining costs of initialization, transport, and retrieval integration.
\end{abstract}


\section{Introduction}

Binary embeddings support similarity search through Hamming distance, which counts the coordinates at which two vectors differ. When embeddings must remain confidential to a computation service, additive homomorphic encryption provides a means of evaluating a packed representation without first decrypting it. Paillier encryption~\cite{paillier} is relevant to this setting because ciphertext multiplication implements plaintext addition. Its cost nevertheless includes large-integer arithmetic, randomized encryption, and the movement of expanded ciphertext representations.

This paper examines the implementation of that arithmetic pipeline in the XTrace SDK. The client packs binary coordinates into separated bit positions, encrypts the packed values, and decodes homomorphically combined ciphertexts into Hamming distances. The implementation includes CPU and GPU backends for both conventional Paillier and a lookup-based configuration. The latter changes the generator, randomization procedure, and decryption exponent in addition to using precomputed tables.

The engineering objective is to reduce repeated work across the complete client call. Message tables replace exponentiation by modular products, cooperative GPU arithmetic distributes each fixed-width integer across a group of threads, and persistent allocations amortize table construction and transfer. The surrounding retrieval code encrypts each query once, decodes returned results in batches, and fetches document contents only after local ranking.

The performance measurements show substantial differences among these implementations. The evaluation fixes the lookup exponent-size parameter at 280 bits and measures large warm batches rather than end-to-end search. The resulting ratios characterize the complete client implementations, including their arithmetic and representation choices. They do not isolate the contribution of each optimization or establish a comparison against the fastest available Paillier implementation.

The study provides the following:
\begin{itemize}[leftmargin=*]
  \item A precise description of the packed Hamming-distance computation, including its carry and plaintext-range conditions.
  \item An account of table-based encryption, cooperative GPU arithmetic, device residency, and retrieval integration.
  \item Measurements of four client configurations, with explicit parameter settings, timing boundaries, and limits on causal interpretation.
\end{itemize}

\paragraph{Implementation Availability.}
An implementation of the Paillier clients is publicly available in the XTrace SDK repository at \url{https://github.com/XTraceAI/xtrace-sdk}. The repository provides installation instructions, usage examples, and tests for readers who wish to evaluate the implementation.

\section{Background and Scope}
\label{sec:background}

\subsection{Cryptographic and Implementation Context}

Paillier's constructions are based on composite residuosity~\cite{paillier}. The earlier Goldwasser--Micali scheme provides probabilistic encryption~\cite{goldwasser-micali}, and Damg{\aa}rd--Jurik extends Paillier to larger plaintext spaces~\cite{damgard-jurik}. These schemes differ in their algebra and ciphertext expansion, so their suitability depends on the computation being implemented.

Precomputation is also an established implementation technique. For example, Rache constructs encryptions using cached ciphertexts and radix decompositions~\cite{rache}. Our message table similarly exploits a fixed radix, but this observation does not establish the security of our randomization procedure or imply equivalence between the constructions. On the arithmetic side, CGBN provides cooperative fixed-width GPU big-integer operations~\cite{cgbn}. The present contribution is the integration and characterization of these techniques for a particular client workload, not a claim that lookup tables or GPU big-integer arithmetic are new.

\subsection{System Model and Information Exposure}

Let \(x,y\in\{0,1\}^d\) be binary embeddings. The desired similarity statistic is the Hamming distance
\[
  \Ham(x,y)=\sum_{i=0}^{d-1} \mathbf{1}[x_i\ne y_i].
\]
The client holds the secret key, and the server combines encrypted query and data vectors under the same public key. The intended confidentiality goal concerns an honest-but-curious server that follows the arithmetic protocol but may inspect its inputs and outputs. We assume that the key-holding client and its device are trusted. Access patterns, record identifiers, candidate counts, and metadata supplied as filters are not hidden.

The returned ciphertexts encrypt packed coordinate sums, not merely the scalar distance. A secret-key holder can recover these sums and, given its query, reconstruct the corresponding data vector. Consequently, the protocol does not provide distance-only disclosure to a mutually untrusted querying client. It also does not establish malicious-server correctness, chosen-ciphertext security, or side-channel resistance. Ranking occurs after decryption at the client, rather than through encrypted top-$k$ selection.

\section{Packed Hamming-Distance Computation}
\label{sec:encoding}

\subsection{Conventional Paillier Baseline}

Let $n=pq$ for distinct primes $p$ and $q$. Conventional Paillier encrypts $m\in\mathbb{Z}_n$ using a suitable public generator $g$ and a fresh random $r\in\mathbb{Z}_n^*$:
\begin{equation}
  \operatorname{Enc}(m,r)=g^m r^n\bmod n^2.
\end{equation}
Its additive homomorphism gives
\[
  \operatorname{Dec}\bigl(\operatorname{Enc}(m_1,r_1)
  \operatorname{Enc}(m_2,r_2)\bmod n^2\bigr)
  =(m_1+m_2)\bmod n.
\]
Our conventional CPU baseline uses $g=1+n$ and decrypts using $\phi(n)=(p-1)(q-1)$ and its inverse modulo $n$. Although the source evaluates $g^m$ with a general modular-exponentiation routine, the binomial theorem gives
\begin{equation}
  (1+n)^m\equiv 1+mn\pmod{n^2}.
  \label{eq:baseline-shortcut}
\end{equation}
Thus a message exponentiation is not inherently necessary for this baseline generator. This distinction matters when interpreting the reported encryption ratios.

\subsection{Encoding and Correctness}

For a chunk of $s$ coordinates, define
\begin{equation}
  E_s(x)=\sum_{i=0}^{s-1}x_i4^{s-1-i},\qquad
  M_s=\sum_{i=0}^{s-1}4^i.
  \label{eq:encoding}
\end{equation}
The big-endian binary representation of $E_s(x)$ is
\[
  0,x_0,0,x_1,\ldots,0,x_{s-1}.
\]
For binary inputs, each base-4 digit of $E_s(x)+E_s(y)$ belongs to $\{0,1,2\}$. No digit therefore carries into its neighbor. The low bit of each digit is one exactly when its two input bits differ. Provided the plaintext sum does not wrap modulo $n$,
\begin{equation}
  \Ham(x,y)=\popcount\bigl((E_s(x)+E_s(y))\mathbin{\&}M_s\bigr),
  \label{eq:decode}
\end{equation}
where $\&$ denotes bitwise AND. These low bits occupy odd zero-based positions when the padded string is read from the left, but even bit offsets when indexed from the least significant bit. Specifying the convention avoids ambiguity in the decode mask.

A sufficient range condition for every pair of $s$-bit chunks is
\begin{equation}
  \frac{2(4^s-1)}{3}<n.
  \label{eq:range}
\end{equation}
This is a correctness requirement, not merely a performance consideration. For longer vectors, the construction applies independently to chunks satisfying this condition and adds their decoded counts. The measured workload has $s=d=512$ and uses one ciphertext per vector. The implementation parameter \texttt{key\_len}=1024 specifies the bit length of each prime, not of $n$. The resulting modulus has 2047 or 2048 bits, comfortably satisfying Equation~\eqref{eq:range} for this workload. No claim about arbitrary chunk sizes follows from this one-chunk experiment.

\section{Lookup Construction and Precomputation}
\label{sec:lookup}

\subsection{Reduced-Exponent Decryption}

The lookup client uses a Paillier-based subgroup construction rather than simply attaching tables to the baseline. Key generation chooses primes $q_1\mid(p-1)$ and $q_2\mid(q-1)$, and sets
\[
  a=\lcm(q_1,q_2).
\]
Generators of orders $q_1$ and $q_2$ modulo $p$ and $q$ are lifted to the squared prime moduli and combined by the Chinese remainder theorem. The intended construction has $\ord_{n^2}(g)=na$, $\ord_n(g)=a$, and
\begin{equation}
  g^a\equiv 1+n\beta\pmod{n^2},\qquad \gcd(\beta,n)=1.
  \label{eq:generator}
\end{equation}
The last condition is necessary for the decryption inverse to exist.

Writing $h=g^n\bmod n^2$, the lookup encryption has the form
\begin{equation}
  c=g^m h^\rho\bmod n^2,
  \label{eq:lookup-encryption}
\end{equation}
where the implemented table sampling determines the effective exponent $\rho$. Since $h^a=1$, exponentiation gives
\[
  c^a\equiv(g^a)^m\equiv 1+nm\beta\pmod{n^2}.
\]
For $u\equiv1\pmod n$, define $L(u)=(u-1)/n$. Decryption therefore recovers
\begin{equation}
  m=L(c^a\bmod n^2)\beta^{-1}\bmod n.
  \label{eq:lookup-decryption}
\end{equation}
The client precomputes $\beta^{-1}$. This derivation establishes the algebraic identity for valid ciphertexts and keys. The reduced exponent lowers the number of wide modular multiplications required during decryption.

\subsection{Parameter Selection}
\label{sec:parameters}

The lookup configuration uses \texttt{alpha\_len}=280, matching the native CUDA extension's default \texttt{ALPHA\_LEN}. Key generation divides this parameter between two 140-bit subgroup primes $q_1$ and $q_2$ and sets $a=\lcm(q_1,q_2)$. For distinct subgroup primes, the resulting exponent has 279 or 280 bits, and the benchmark records its actual bit length. The parameter specifies the target size of the decryption exponent, not the modulus size or a numerical security level.

The CPU and GPU lookup configurations use this same parameter setting. It should be supplied explicitly when reproducing the study, since defaults at the Python and native interfaces need not coincide. The present paper evaluates implementation performance at these fixed parameters. It does not derive a new security proof or infer cryptographic security from timing measurements.

\subsection{Message Lookup Table}

For the nonstandard generator in Equation~\eqref{eq:lookup-encryption}, the client decomposes a plaintext into base-256 digits:
\[
  m=\sum_{i=0}^{b-1}d_i2^{8i},\qquad d_i\in\{0,\ldots,255\}.
\]
It precomputes $T_{i,j}=g^{j2^{8i}}\bmod n^2$ for each digit position and byte value. Online encryption then evaluates
\[
  g^m\equiv\prod_{i=0}^{b-1}T_{i,d_i}\pmod{n^2}.
\]
The table depends only on public parameters and can be shared between compatible contexts. At the configured capacity of 2048 plaintext bits, $b=256$. With 4096-bit fixed-width entries, the $256\times256$ message table occupies 32\,MiB before allocator or host-object overhead. This is a representation-size calculation, not a measured process-memory result.

\subsection{Precomputed Randomized Masking}

The random factor is also precomputed. The client constructs 256 entries
\[
  \eta_j=h^{r_j}\bmod n^2
\]
and multiplies 14 independently selected entries, with replacement, for each encryption. The resulting factor is $h^{\sum_{t=1}^{14}r_{J_t}}$, so it preserves the algebraic decryption identity. Noise tables are retained per client instance rather than shared through the deterministic message-table cache.

Noise-table state is distinct from the deterministic public message table. The implementation maintains this distinction in its caching policy and samples fresh indices for each ciphertext. Analysis of the resulting randomization distribution and its reuse is separate from the algebraic correctness and performance evaluation presented here.

\section{GPU Arithmetic Implementation}

\subsection{Cooperative Fixed-Width Arithmetic}

The GPU implementation uses CUDA and CGBN~\cite{cgbn}. Each CGBN environment has a fixed integer width, with a cooperative group operating on one big-integer instance. The lookup extension's default configuration assigns 32 threads per instance and reserves 4096 bits for residues modulo $n^2$ when \texttt{KEY\_BITS}=1024. Wider intermediate products support modular reduction without truncating the arithmetic result.

Independent ciphertext chunks supply parallel work across cooperative groups. Table-based encryption consists primarily of indexed loads and modular products, while decryption performs exponentiation followed by the $L$ map and multiplication by a precomputed inverse. Batching amortizes kernel-launch and host/device transfer costs. The measurements in Section~\ref{sec:evaluation} characterize a batch of 10,000 vectors and do not establish a crossover point at which GPU execution becomes preferable.

\subsection{Combination, Decode, and Re-encryption}

Homomorphic combination is pointwise ciphertext multiplication modulo $n^2$. After decryption, the GPU decode path applies the alternating-bit mask in Equation~\eqref{eq:decode} and counts the selected bits. Keeping this operation on the device avoids returning full plaintext chunks merely to compute a scalar count. The server multiplication and client decoding are distinct operations, and only the latter is included in the decode measurement.

The lookup extension also implements a fused decrypt-then-re-encrypt kernel. This operation retains intermediate plaintext on the device and encrypts it under a target key, provided the plaintext fits the target modulus. It requires access to the source secret key and is therefore a trusted re-encryption operation, not a proxy re-encryption protocol. It is not included in the four-configuration evaluation, and no quantitative speedup is claimed for it here.

\section{Runtime and Retrieval-System Optimizations}

\subsection{Lazy Initialization and Persistent Tables}

The lookup GPU backend separates context metadata loading from expensive runtime preparation. Configuration loading can defer table construction and device allocation until the tables are needed. A process-wide host cache reuses deterministic message tables for compatible public parameters. Lookup and noise tables remain resident on the GPU across calls, subject to the lifetime of the client context. The CPU lookup path need not follow the same initialization policy. Warm-state measurements therefore exclude costs that remain relevant to newly created contexts and short-lived applications.

\subsection{Raw-Byte Interchange}

The integration supports ciphertexts represented as little-endian raw bytes. JSON requests encode these bytes using base64, while MessagePack responses can carry binary values without that textual expansion. The Python client accepts byte-valued ciphertexts and normalizes them as required by the selected backend.

Representation is not uniform across all implementations. Native bindings and compatibility paths also expose Python integers, GMP-backed objects, and string conversions. In particular, the conventional GPU wrapper retains hexadecimal conversion in its decode interface. Byte-oriented transport should therefore not be conflated with a claim that every measured Python/C++ call is conversion-free or zero-copy. The reported public-method timings include the conversions performed by each measured backend.

\subsection{Retrieval Orchestration}

The SDK retriever converts a query embedding to binary form, encrypts it once, and submits it with an execution-context identifier. Optional metadata and range filters are passed in the same request to restrict candidate evaluation. Returned encrypted results are associated with record identifiers. By default, the retriever decodes them through \texttt{decode\_hamming\_client\_batch}.

An explicitly selected multiprocessing mode instead dispatches individual decode calls through a process pool. This is an alternative execution mode, not an automatic fallback or an additional layer of GPU parallelism. Its benefit depends on process startup, serialization, and backend compatibility. Ranking uses NumPy sorting, and content retrieval and AES decryption are deferred until the top-$k$ identifiers have been selected. These observations concern SDK orchestration of server requests. They do not demonstrate server-internal parallelism, and neither server scalability nor multiprocessing speedup is measured here.

\section{Evaluation}
\label{sec:evaluation}

\subsection{Methodology}

We benchmarked the public batch methods for vector encryption and Hamming-distance decoding:
\[
  \texttt{encrypt\_vec\_batch}
  \quad\text{and}\quad
  \texttt{decode\_hamming\_client\_batch}.
\]
The benchmark used $N=10{,}000$ random binary embeddings of dimension $d=512$ and one random binary query. The input generator used seed 1337. Each of four configurations constructed a separate key pair with \texttt{key\_len}=1024. Both lookup clients were explicitly configured with \texttt{alpha\_len}=280, corresponding to Section~\ref{sec:parameters}. The harness verified the selected CPU or GPU backend and the runtime exponent-size parameter before measurement. The configurations are conventional Paillier CPU, lookup CPU, conventional Paillier GPU, and lookup GPU. The term ``lookup'' denotes the complete construction in Section~\ref{sec:lookup}, not an isolated table toggle.

Each client first processed a warmup batch of 128 vectors through encryption and distance decoding. The experiment then measured \BenchmarkRepeats{} encryption calls followed by \BenchmarkRepeats{} decode calls per configuration, using the same input batch and key within that configuration. We report the median batch time for each method. The harness measured wall-clock time with Python's \texttt{time.perf\_counter}, with garbage collection before each timed call. Calls returned host-visible results, so these measurements include the corresponding wrapper work and device transfers, rather than only CUDA kernel time. The CPU batch methods process vectors serially. The optional retriever process pool was not used.

Key generation, initialization, warmup, query encryption, and construction of the combined ciphertexts were outside the timed regions. After the encryption trials, the harness formed the decode batch from the final encryption output by multiplying each ciphertext with an encrypted query modulo $n^2$. This batch was reused across the decode trials. The harness did not contact a server. All 10,000 returned distances in every decode trial matched the plaintext reference, for 120,000 checked results across the four configurations. These checks establish agreement on the sampled workload, not exhaustive correctness or cryptographic security.

Measurements were collected on an AMD Ryzen 7 5800X CPU and an NVIDIA GeForce RTX 3080 with 10\,GB of memory, using Python 3.12.3, gmpy2 2.3.0, GMP 6.3.0, and NVIDIA driver 580.173.02. The archived record includes individual timings, runtime configurations, the SDK source revision, installed toolchain information, and hashes of the relevant sources and native extensions. The experiment used a shared workstation without exclusive GPU access. Three trials on one key per configuration support descriptive comparisons, not confidence intervals or estimates of variation across independently generated keys.

\subsection{Results}

For batch time $T$ in seconds, the amortized cost is $1000T/N$ milliseconds per vector and throughput is $N/T$ vectors per second. The former is not the response latency of an isolated vector or a complete search request.

\begin{table}[!htbp]
\centering
\small
\setlength{\tabcolsep}{5pt}
\begin{tabular}{lrrrr}
\toprule
& \multicolumn{2}{c}{Amortized cost (ms/vector)} & \multicolumn{2}{c}{Throughput (vectors/s)} \\
\cmidrule(lr){2-3}\cmidrule(lr){4-5}
Client & Encrypt & Decode & Encrypt & Decode \\
\midrule
Paillier CPU & \BaselineCPUEncryptMs & \BaselineCPUDecodeMs & \BaselineCPUEncryptRate & \BaselineCPUDecodeRate \\
Lookup CPU & \LookupCPUEncryptMs & \LookupCPUDecodeMs & \LookupCPUEncryptRate & \LookupCPUDecodeRate \\
Paillier GPU & \BaselineGPUEncryptMs & \BaselineGPUDecodeMs & \BaselineGPUEncryptRate & \BaselineGPUDecodeRate \\
Lookup GPU & \LookupGPUEncryptMs & \LookupGPUDecodeMs & \LookupGPUEncryptRate & \LookupGPUDecodeRate \\
\bottomrule
\end{tabular}
\caption{Warm-batch measurements for 10,000 random 512-bit embeddings, computed from the median of \BenchmarkRepeats{} trials per method and configuration. Both lookup clients use \texttt{alpha\_len}=280. Costs are amortized per vector.}
\label{tab:results}
\end{table}

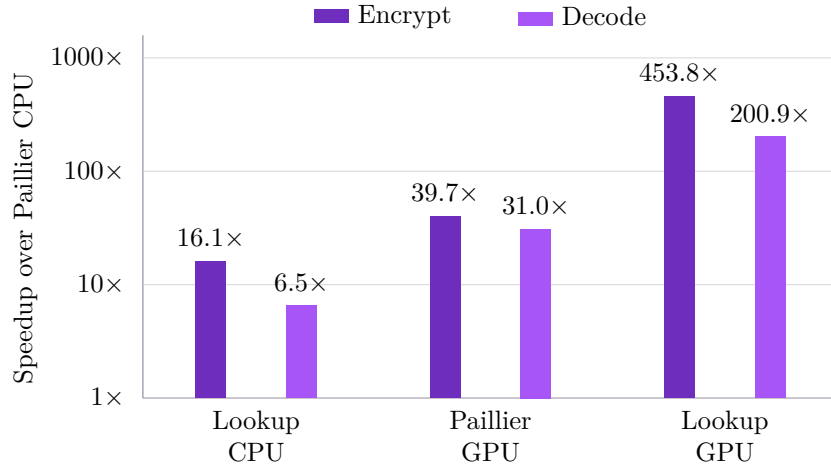
\begin{figure}[!htbp]
\centering
\begin{tikzpicture}[x=1cm,y=1cm]
\small
\draw[xtraceInk!55, line width=0.4pt] (0,0) -- (9.2,0);
\draw[xtraceInk!55, line width=0.4pt] (0,0) -- (0,4.8);
\foreach \y/\tick in {0/1,1.5/10,3.0/100,4.5/1000} {
  \draw[xtraceInk!18, line width=0.3pt] (0,\y) -- (9.2,\y);
  \node[anchor=east] at (-0.12,\y) {$\tick\times$};
}
\node[rotate=90, anchor=center] at (-1.55,2.25) {Speedup over Paillier CPU};
\node[anchor=west] at (2.15,5.05) {\textcolor{xtracePurple}{\rule{0.45cm}{0.18cm}} Encrypt};
\node[anchor=west] at (4.85,5.05) {\textcolor{xtraceViolet}{\rule{0.45cm}{0.18cm}} Decode};
\foreach \x/\ratio/\shade in {
  0.9/\LookupCPUEncryptSpeedup/xtracePurple,2.1/\LookupCPUDecodeSpeedup/xtraceViolet,
  4.0/\BaselineGPUEncryptSpeedup/xtracePurple,5.2/\BaselineGPUDecodeSpeedup/xtraceViolet,
  7.1/\LookupGPUEncryptSpeedup/xtracePurple,8.3/\LookupGPUDecodeSpeedup/xtraceViolet} {
  \pgfmathsetmacro{\height}{1.5*ln(\ratio)/ln(10)}
  \fill[\shade] (\x-0.2,0) rectangle (\x+0.2,\height);
  \node[anchor=south] at (\x,\height+0.05) {$\ratio\times$};
}
\node[align=center] at (1.5,-0.5) {Lookup\\CPU};
\node[align=center] at (4.6,-0.5) {Paillier\\GPU};
\node[align=center] at (7.7,-0.5) {Lookup\\GPU};
\end{tikzpicture}
\caption{Speedup relative to the measured Paillier CPU implementation on a logarithmic axis, using ratios of median batch times. Ratios compare complete client configurations. The lookup variants combine table-based encryption and reduced-exponent decryption, so the comparison is not an isolated lookup-table ablation.}
\label{fig:speedup}
\end{figure}

Table~\ref{tab:results} reports the normalized costs and throughputs, and Figure~\ref{fig:speedup} reports the ratios. Relative to the measured CPU baseline, lookup CPU is $\LookupCPUEncryptSpeedup\times$ faster for encryption and $\LookupCPUDecodeSpeedup\times$ faster for decode. Conventional Paillier GPU has ratios of $\BaselineGPUEncryptSpeedup\times$ and $\BaselineGPUDecodeSpeedup\times$, respectively. Lookup GPU has the largest ratios, $\LookupGPUEncryptSpeedup\times$ and $\LookupGPUDecodeSpeedup\times$. Within the lookup family, the GPU implementation is $\LookupGPUOverCPUEncrypt\times$ faster for encryption and $\LookupGPUOverCPUDecode\times$ faster for decode than its CPU counterpart. Appendix~\ref{app:timing-ranges} reports the observed timing ranges.

\subsection{Interpretation and Experimental Limits}

The CPU/GPU comparisons within each construction provide evidence that batched accelerator execution benefits this implementation. They do not isolate GPU arithmetic from differences in wrappers, serialization, or decoding. Likewise, lookup/non-lookup comparisons jointly change the generator, exponent length, randomization, and table usage. In particular, faster lookup decoding cannot be attributed to the message table, which is not used by the decryption formula. The experiment is not an ablation study of individual optimizations.

The conventional CPU reference also leaves Equation~\eqref{eq:baseline-shortcut} unused and does not provide a tuned multicore or CRT-based decryption baseline. The headline ratios are therefore relative to this specific reference, not to the fastest available Paillier software. A broader evaluation would add tuned CPU baselines, repeated measurements across multiple keys, and experiments varying batch size, embedding dimension, and hardware. Cold-start cost, peak memory, network transfer, and full retrieval latency would require separate measurements.

\section{Explored Alternatives}

During development, we explored CPU vectorization through Intel IPP Cryptography and the \texttt{ippcp.h} interface. Intel's Cryptography Primitives library supports several SIMD instruction sets, including AVX-family extensions~\cite{intel-crypto-primitives}. Our prototype did not outperform the GMP-based path. Data rearrangement, carry handling, and movement between vector operations are possible explanations, but we do not report profiling measurements that isolate these mechanisms. This observation is not a general conclusion about AVX or Intel's library.

We also implemented clients for Goldwasser--Micali and Damg{\aa}rd--Jurik encryption~\cite{goldwasser-micali,damgard-jurik}. Neither prototype provided higher performance than our Paillier implementation in the development workloads we tried. These exploratory comparisons are not accompanied by a controlled benchmark record in this paper. They explain the engineering direction taken, but do not establish superiority over those schemes at matched security or across other workloads.

\section{Limitations and Future Work}

The evaluation concerns client-side computation at fixed cryptographic parameters. It does not measure complete retrieval latency or establish guarantees beyond the system model in Section~\ref{sec:background}. Metadata leakage, disclosure of packed sums to the key-holding client, and result integrity remain separate protocol considerations.

The systems techniques have separate limitations. Fixed-width kernels constrain supported parameters, precomputation consumes memory, and persistent tables favor long-lived contexts. Small batches and cold starts may have different performance characteristics. The current evaluation does not quantify these tradeoffs or the benefit of fused re-encryption and retrieval parallelism.

Operational experience motivated attention to ciphertext transmission after local arithmetic became faster. With a 4096-bit ciphertext representation, 10,000 one-chunk results occupy 5.12\,MB before identifiers and protocol framing. This is a payload-size calculation, not a measured network bottleneck in the experiment. Byte-oriented transport can remove textual overhead but does not remove the cryptosystem's ciphertext expansion. Future work should measure end-to-end transfer costs and investigate candidate pruning, request batching, and protocols that return fewer encrypted values. Reducing protocol traffic is a more defensible objective than assuming that pseudorandom ciphertext bytes admit substantial general-purpose compression.

\section{Conclusion}

We described a batched Paillier-based Hamming-distance client and the interactions among packing, precomputation, GPU arithmetic, and runtime state. Both GPU configurations outperformed their CPU counterparts on the evaluated batch. With the 280-bit exponent-size parameter, the lookup GPU configuration achieved \LookupGPUEncryptRateRounded{} encryptions/s and \LookupGPUDecodeRateRounded{} Hamming-distance decodes/s, based on median batch times. Relative to the measured conventional CPU baseline, the corresponding speedups were $\LookupGPUEncryptSpeedup\times$ and $\LookupGPUDecodeSpeedup\times$. These results characterize the combined implementation rather than individual optimizations in isolation. Further evaluation should extend the comparison to tuned CPU baselines, additional workloads, and complete retrieval measurements, with particular attention to ciphertext transport.

\appendix
\section{Timing Ranges and Reproducibility}
\label{app:timing-ranges}

Table~\ref{tab:ranges} reports the minimum and maximum amortized costs observed in the \BenchmarkRepeats{} trials for each method. These ranges describe variability within this run and are not confidence intervals. The ancillary materials contain the raw JSON record, a snapshot of the benchmark harness, and the script that generates the numerical values used throughout the manuscript. The harness requires the corresponding SDK and native extensions. Artifact hashes identify the measured files but do not replace the implementation dependencies needed for a complete reproduction.

\begin{table}[!htbp]
\centering
\small
\begin{tabular}{lrr}
\toprule
Client & Encrypt range (ms/vector) & Decode range (ms/vector) \\
\midrule
Paillier CPU & \BaselineCPUEncryptMinMs--\BaselineCPUEncryptMaxMs & \BaselineCPUDecodeMinMs--\BaselineCPUDecodeMaxMs \\
Lookup CPU & \LookupCPUEncryptMinMs--\LookupCPUEncryptMaxMs & \LookupCPUDecodeMinMs--\LookupCPUDecodeMaxMs \\
Paillier GPU & \BaselineGPUEncryptMinMs--\BaselineGPUEncryptMaxMs & \BaselineGPUDecodeMinMs--\BaselineGPUDecodeMaxMs \\
Lookup GPU & \LookupGPUEncryptMinMs--\LookupGPUEncryptMaxMs & \LookupGPUDecodeMinMs--\LookupGPUDecodeMaxMs \\
\bottomrule
\end{tabular}
\caption{Observed minimum and maximum amortized costs across \BenchmarkRepeats{} warm-batch trials per method and configuration.}
\label{tab:ranges}
\end{table}


\begin{thebibliography}{9}
\raggedright
\small

\bibitem{paillier}
P. Paillier.
\newblock Public-key cryptosystems based on composite degree residuosity classes.
\newblock In \emph{Advances in Cryptology, EUROCRYPT 1999}, LNCS 1592, pp.~223--238. Springer, 1999.
\newblock \href{https://doi.org/10.1007/3-540-48910-X_16}{doi:10.1007/3-540-48910-X\_16}.

\bibitem{goldwasser-micali}
S. Goldwasser and S. Micali.
\newblock Probabilistic encryption.
\newblock \emph{Journal of Computer and System Sciences}, 28(2):270--299, 1984.
\newblock \href{https://doi.org/10.1016/0022-0000(84)90070-9}{doi:10.1016/0022-0000(84)90070-9}.

\bibitem{damgard-jurik}
I. Damg{\aa}rd and M. Jurik.
\newblock A generalisation, a simplification and some applications of Paillier's probabilistic public-key system.
\newblock In \emph{Public Key Cryptography, PKC 2001}, pp.~119--136. Springer, 2001.
\newblock \href{https://doi.org/10.1007/3-540-44586-2_9}{doi:10.1007/3-540-44586-2\_9}.

\bibitem{rache}
D. Zhao.
\newblock Rache: Radix-additive caching for homomorphic encryption.
\newblock arXiv:2201.04255, 2022.
\newblock \url{https://arxiv.org/abs/2201.04255}.

\bibitem{cgbn}
NVIDIA Research.
\newblock CGBN: CUDA accelerated multiple precision arithmetic using cooperative groups.
\newblock \url{https://github.com/NVlabs/CGBN}.
\newblock Accessed September 15, 2026.

\bibitem{intel-crypto-primitives}
Intel.
\newblock Intel Cryptography Primitives Library.
\newblock \url{https://github.com/intel/cryptography-primitives}.
\newblock Accessed September 15, 2026.

\end{thebibliography}
\end{document}